# Controlling Light with Nonlinear Quasicrystal Metasurfaces

Yutao Tang,[1,2,3,†] Junhong Deng,[1,2,†] King Fai Li,[1,2,†] Mingke Jin,[1] Jack Ng,[3] Guixin Li[1,2,*]

[1]Department of Materials Science and Engineering, Southern University of Science and Technology, Shenzhen, 518055 China.

[2]Shenzhen Institute for Quantum Science and Engineering, Southern University of Science and Technology, Shenzhen, 518055 China.

[3]Department of Physics and Institute of Computational and Theoretical Studies, Hong Kong Baptist University, Hong Kong, China.

Metasurface, a kind of two-dimensional structured medium, represents a novel platform to manipulate the propagation of light at subwavelength scale. In linear optical regime, many interesting topics such as planar metalens, metasurface optical holography and so on have been widely investigated. Recently, metasurfaces go into nonlinear optical regime. While it is recognized that the local symmetry of the meta-atoms plays vital roles, its relationship with global symmetry of the nonlinear metasurfaces remains elusive. According to the Penrose tiling and the newly proposed hexagonal quasicrystalline tiling, here we designed and fabricated the nonlinear optical quasicrystal metasurfaces based on the geometric phase controlled plasmonic meta-atoms with local rotational symmetry. The second harmonic waves will be determined by both the tiling schemes of quasicrystal metasurfaces and the local symmetry of meta-atoms they consist of. The proposed concept opens new routes for designing nonlinear metasurface crystals with desired optical functionalities.

Metasurfaces, as the two-dimensional (2D) artificially engineered surfaces, have enabled abundant optical functionalities and hold great potentials for applications in integrated optics thanks to their compactness and flexibilities of design [1–4]. In the past years, photonic metasurfaces, which outperform traditional bulky optical components in many aspects, have been extensively explored in areas ranging from planar lens [5–8] to optical holography [9–11] and so on. Apart from the practical applications of metasurfaces, novel optical physics, to name a few, optical geometric Pancharatnam-Berry (P-B) phase [12–14], generalized Snell's law [1], spin-orbit interaction [15–18], quantum information processing [19,20] and so on, also attract extensive attentions. In the context of linear optics, how to elaborately order the meta-atoms in metasurfaces is important for realizing specific optical functions such as focusing, beam-splitting, generation of optical vortex, etc. In addition to the periodic lattices, the adoption of quasicrystalline structure [21–24] may also bring new degrees of freedom to the optical functions of metasurfaces. For example, optical spin Hall effect has been observed on Penrose-type plasmonic metasurface [25].

Recently, the research of optical metasurface goes into nonlinear optical regime [26–28]. The key issues in this field mainly involve how to control the amplitude, phase and polarization of light in nonlinear optical processes. Strong localization of electromagnetic waves in vicinity of the metasurface can be used to boost the efficiency of second harmonic generation (SHG) [29–33], third harmonic generation (THG) [34,35] and four-wave mixing (FWM) [36], etc. In addition, the active medium that forms the plasmonic [30–32,36], metal-dielectric hybrid [29,34,35], or dielectric [37,38] meta-atoms are also important for certain nonlinear optical processes. There has also been strong interests to locally control the phase of nonlinear waves by using electric polling method, but only until recently, they can be continuously manipulated with the help of nonlinear geometric P-B phase of meta-atoms [28,35]. Apart from this, the nonlinear polarizations can also be well controlled by the local symmetry of meta-atom and the polarization of fundamental wave (FW). It has been verified that nonlinear optical metasurfaces has many applications in fields such as imaging, optical encryption and holography and so on [28]. Now it is timely to pay more attentions to the relationship between local and global symmetries of metasurfaces, which may introduce new degrees of freedom to control the nonlinear optical radiation. As we all know, as an ordered state of matter lack of periodicity, quasicrystal has revealed rich physics and has gained great success especially after the discovery of natural quasicrystal [21–24]. As shown in Fig. 1, we take the advantages of both the nonlinear P-B phase controlled meta-atoms and various quasicrystal tiling schemes to construct the nonlinear optical quasicrystal metasurface (NOQCM). A family of meta-atoms enabled by nonlinear geometric P-B phase can be assembled together to form the NOQCM, of which the nonlinear radiation including SHG, THG, and FWM is governed by both local and global symmetries of the meta-atoms. In comparison, the local symmetry in the nonlinear photonic quasicrystal is quite difficult be engineered at will [39].

Without lack of generalization, we designed NOQCM samples according to the famous Penrose tiling, and the newly proposed hexagonal quasicrystalline (HQC) tiling [40]. Compared to the extensively investigated Penrose structures, the HQC metasurface we designed is the first experimental demonstration of its kind. The measured SHG from the metasurfaces show strong spin-dependent optical properties, while the diffraction patterns reveal the unique rotational symmetries of both the meta-atoms and the quasiperiodic crystal they form. We believe that the novel nonlinear optical quasicrystal metasurfaces provide a novel platform for studying the rich physics of nonlinear optics, quasicrystalline ordered state, and topological photonics and so on.

According to the crystallographic restriction theorem, the possible rotational symmetry of a crystal is strictly limited to two-, three-, four- and six-fold. The discovery of ten-fold X-ray diffraction pattern in 1982 and the subsequent introduction of the concept of quasicrystal completely revolutionizes the definition of crystal [21–24]. Since then quite a lot of quasicrystal designs were proposed and experimentally demonstrated. In Penrose tiling scheme [Fig. 1(c)] hides the gold mean ratio $(1+\sqrt{5})/2$, which is closely related to the Fibonacci sequence generated by a simple substitution rule. The generalized substitution rule also leads to other metallic mean ratios associated to different quasicrystal tiling accordingly [41]. The second-order substitution rule gives the silver mean, and the corresponding tiling has an eight-fold rotational symmetry. As is shown in Fig. 1(d-e) and Fig. S1, the third-order tiling corresponds to the bronze-mean and has a six-fold rotational symmetry which usually happens in conventional crystal based on the crystallographic restriction theorem.

The nonlinear meta-atoms used to construct the NOQCM is a gold nanostructure with three-fold (C3) rotational symmetry. As the inversion symmetry is broken in the C3 meta-atom, SHG is allowed and could be greatly enhanced when the localized plasmon resonance of the fundamental wave (FW) is excited. From the symmetry selection rule of nonlinear harmonic generations, it is known that the local rotational symmetry of the meta-atom selectively allows certain order of harmonic generation under normal incidence of the FW. For example, the plasmonic meta-atoms with C3 rotational symmetry emit circularly polarized SHG wave that has handiness opposite to that of the FW. More interestingly, a nonlinear optical P-B phase $3\sigma\theta$ of the C3 meta-atom is introduced to the effective second-order susceptibility $\chi^{(2)}$, where $\theta$ is the in-plane orientation angle of the meta-atom; $\sigma = \pm 1$ represents the left- and right-circular polarizations (LCP and RCP) of the FW, respectively.

The quasicrystal metasurfaces can be artificially decorated using nonlinear optical meta-atoms with various local symmetries. In the Penrose case, we set the side length of the rhombuses tiles as 800 nm and allocate the C3 gold meta-atoms at each lattice point with unidirectional orientation [Fig. 1(a)]. In HQC case, there are four kinds of quasicrystal lattice in general [41], corresponding to two distinct assembly rules applied to two different basic dodecagons [Fig. S1]. Among these four types of quasicrystal lattice, only one possesses the self-similarity. The four kinds of lattice can be divided into sparse and dense versions ascribed to the assembly rules they obey. The elements of HQC are three tiles, small and large equal-lateral triangles (ST and LT), and rectangles. In our design, the side length of the small triangles is set as 800 nm, and the side length of the large triangles can be derived from the self-similarity condition of the subdivision process [41]. The small and large triangles in the HQC tiling provide great convenience to arrange the C3 meta-atoms. We can arrange the meta-atoms at the center of each triangle and align their arms in two ways: one has broken inversion symmetry [Fig. 1(c)] and another one has inversion symmetry [Fig. 1(d)]. The NOQCMs are fabricated by using standard electron beam lithography followed by the metal lift-off process [41]. A 30 nm thick gold metasurfaces are sitting on top of an indium tin oxide coated glass substrate, with the width and length of the arms of C3 meta-atoms are 80 nm and 180 nm, respectively [Fig. 2(a-c) and Fig. S2].

Next, the linear optical property of the quasicrystal metasurfaces are characterized using the homemade transmission setup for infrared and visible regimes [Fig. 2(d-f), Figs. S3 and S4]. The measurements are conducted under the illumination of horizontally polarized (H-) white light, and both horizontally (H-H) and vertically (H-V) polarized transmission spectra are recorded. The near-to-zero H-V transmittance indicates that there is almost no polarization conversion in the linear optical regime. For H-H spectra, all metasurfaces exhibit a broadband transmission valley spanning from 1050 nm to 1600 nm, which are mainly due to the localized plasmon resonances of the gold meta-atoms. As the optical axis of meta-atoms is parallel to the propagation direction of light and the coupling between the meta-atoms is very weak in all NOQCMs [Fig. 2, (e) and (f)], thus the transmission spectra are insensitive to the polarization of FW and the lattice structures.

Then, the SHG responses of the quasicrystal metasurfaces are measured by using a femtosecond-laser pumped optical parametric oscillator (pumping wavelength: 820 nm; repetition frequency: 80 MHz). The circularly polarized femtosecond FW is focused onto the metasurfaces with a spot size about 20 μm in diameter. A pair of circular polarizers are used to select the four combinations of spin states of FW and SHG waves. We choose the Penrose and HQC metasurfaces consisting of meta-atoms with unidirectional orientations as illustrative examples. In all the nonlinear optical measurements SHG intensities with circular polarizations opposite to that of FW (LCP-RCP and RCP-LCP) are much stronger than for the cases that have the same handiness (LCP-LCP and RCP-RCP). The resonant peaks of SHG responses are observed at wavelength 1275 nm [Fig. 3(a)] and 1250 nm [Fig. 3(d)] for the Penrose and HQC metasurfaces, respectively. The measured results agree well with the symmetry selection rule for SHG on C3 meta-atoms. As is shown in Fig. 3(c) and 3(f), we further measured the power dependences of SHG waves from the nonlinear optical quasicrystal metasurfaces at resonant pumping wavelengths of 1275 nm and 1250 nm.

The values of slope in the log-log scale are near to two, indicating a second-order nonlinear optical process. In addition, the SHG spectra for the four kinds of polarization measurement schemes are shown in Fig. 3(c) and 3(f). Compared to the strong SHG radiation from the quasicrystal metasurface with unidirectional orientations of meta-atoms, the measured SHG from symmetric HQC samples are much weaker [Fig. S5].

The optical diffraction technique is also exploited to investigate the nonlinear optical quasicrystal metasurfaces in both linear [Fig. S6] and nonlinear [Fig. 4] optical regimes. The nonlinear optical diffraction from the Penrose and HQC metasurfaces are also measured at the resonant fundamental wavelengths of 1275 nm and 1250 nm, respectively. Based on the spin dependent SHG responses in Fig. 3, we measure the nonlinear diffraction patterns under various polarization combinations. The SHG radiation patterns are collected by using the objective lens (NA=0.25) and imaged in the $k$-space [Fig. 4(a-c) and Fig. S7], where $k_0$ is the vacuum wavevector of SHG wave, $k_x$ and $k_y$ are the reciprocal lattice vectors along $x$- and $y$-axis. It is found that the ten-fold and six-fold rotational symmetries of optical diffraction pattern also exist in the nonlinear optical regime. However, a phenomenon worth to be mentioned is that the SHG diffraction patterns of the symmetric HQC metasurfaces [Fig. 4(b) and Fig. S7(e-h)] exhibit significant peripheral spots, but their $0^{th}$ orders disappear. This is consistent with previous observations that the SHG efficiency of symmetric metasurfaces are much lower than that of the asymmetric ones [Fig. S5].

To better understand the behavior of the SHG radiations from the quasicrystal metasurfaces, we develop the dipolar radiation model in both linear and nonlinear optical regimes. For linear optical diffraction, the electric fields of the dipoles that represent the meta-atoms are assumed to have relative phase of zero. In comparison, when excited by a circularly FW with helicity $\sigma = \pm 1$, the SHG wave with opposite helicity $-\sigma$ experiences a nonlinear geometric phase of $\exp(i3\sigma\theta)$, where $\theta$ is the in-plane orientation angle of the C3 meta-atoms. The far field radiation of both linear and SHG waves are calculated by using Fourier transformation method [41]. The results for linear optical calculations [Fig. S6(d-f)] and the nonlinear optical calculations [Fig. 4(d-f) and Fig. S8] perfectly agree with the measured ones. After reviewing the distribution of the meta-atoms in the symmetric HQC metasurfaces, we can find that two neighboring meta-atoms [neighboring STs or LTs in Fig. 1(c)] have a rotation angle difference of $\pi/3$. Therefore, the SHG signals generated by any pair of two adjacent meta-atoms will have a $\pi$ phase difference in the far field, leading to a destructive interference of SHG waves at the 0$^{th}$ diffraction order.

In summary, we have designed and fabricated nonlinear optical quasicrystal metasurfaces by assembling the geometric P-B phase controlled plasmonic meta-atoms into the macroscopic quasicrystal lattices. From the systematic study of optical behaviors of the quasicrystal metasurfaces, we successfully demonstrate that the nonlinear SHG radiation can be manipulated by both the local and global symmetries of the metasurface crystals. The theoretical model, developed for interpreting the far field nonlinear optical radiations, also enables us with powerful tool for predicting the functionalities of other quasicrystal metasurfaces. It is anticipated that the concept of quasicrystal metasurfaces not only represent novel strategy for controlling various nonlinear optical processes at the subwavelength scale, but also offers an integrated platform for quantum information processing, optical modulation, all-optical optical switching and so on.

**Acknowledgement**

This research was supported by the Guangdong Provincial Innovation and Entrepreneurship Project (2017ZT07C071), Applied Science and Technology Project of Guangdong Science and Technology Department (2017B090918001), National Natural Science Foundation of China (11774145) and Natural Science Foundation of Shenzhen Innovation Committee (JCYJ20170412153113701).

† These authors contribute equal to this work.

*ligx@sustc.edu.cn

**Figures and Captions**

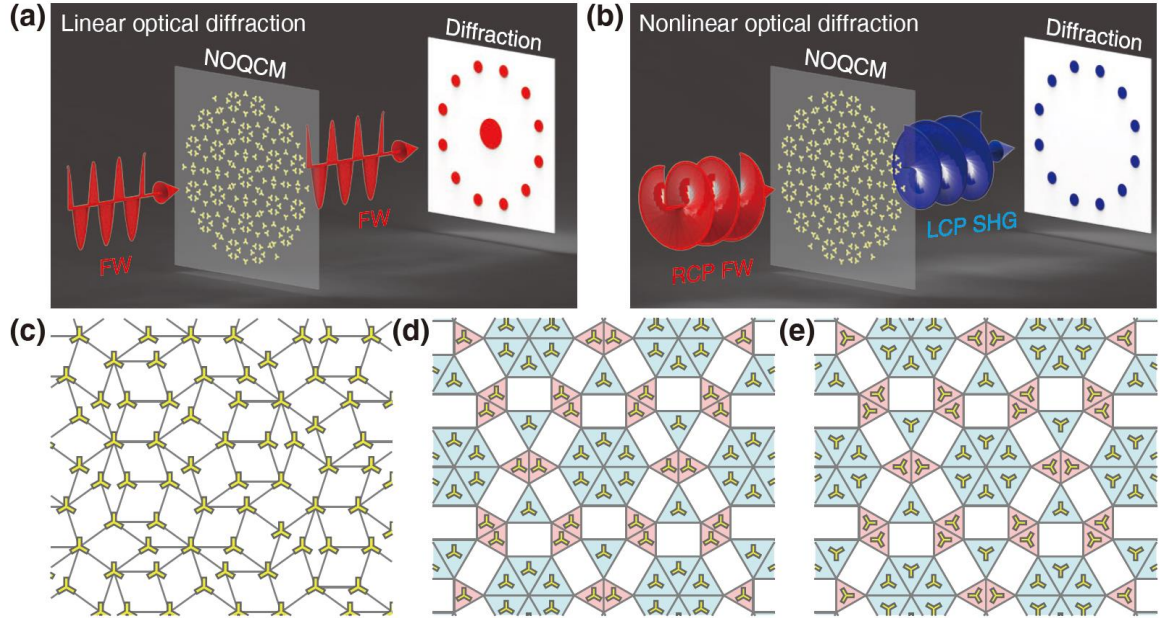

FIG. 1. Demonstrations and designs of nonlinear optical quasicrystal metasurfaces (NOQCMs) consisting of plasmonic meta-atoms with C3 rotational symmetry. (a) Linear optical diffraction of a NOQCM. (b) Nonlinear optical diffraction of the same NOQCM as in (a). (c) Penrose type quasicrystal metasurface with five-fold rotational symmetry. The grid depicts the Penrose tiling with thin and flat rhombuses of 800 nm side length. The meta-atoms are oriented to the same direction. (d) Hexagonal quasicrystalline (HQC) type quasicrystal metasurface. The HQC metasurface consists of small and large equilateral triangles and complementary rectangles. The meta-atoms are aligned to the same direction. (e) Symmetric design of HQC metasurface, where the meta-atoms together with the occupied triangles have C3 rotational symmetry.

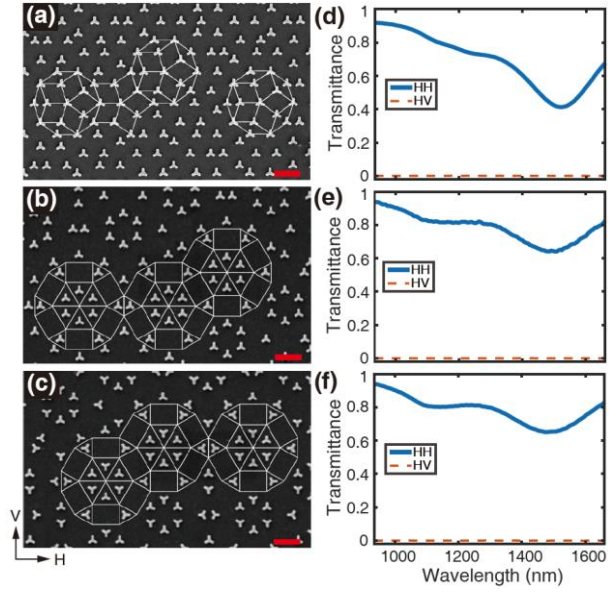

FIG. 2. Scanning electron microscope (SEM) images and measured transmission spectra of the quasicrystal metasurfaces. (a-c) SEM images of the metasurfaces corresponding to the designs in Fig. 1. Scale bars: 1 μm. (d-f) Measured near infrared transmission spectra of the metasurfaces shown left. The incident light is horizontally polarized and both horizontal (HH) and vertical (HV) polarized transmission spectra are measured. The resonant transmission valleys are consistently located at the wavelength of about 1520 nm, and all spectra are similar in shapes owing to the same localized plasmon resonance of the C3 gold meta-atoms.

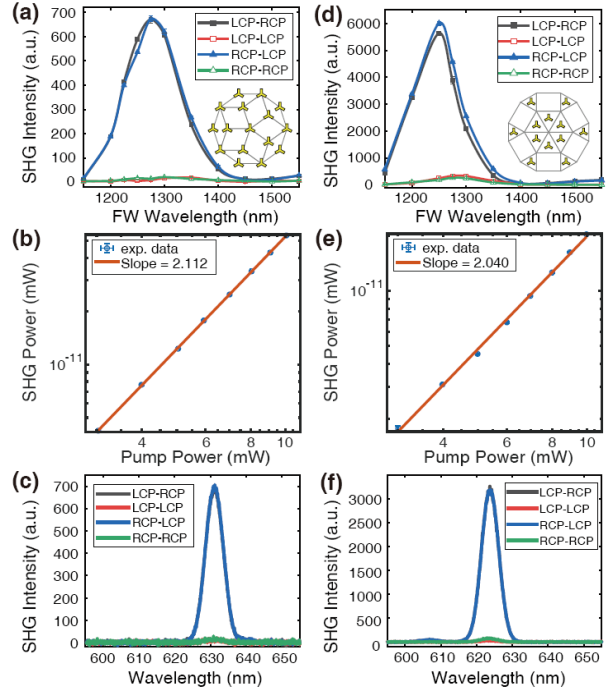

FIG. 3. Measured nonlinear optical properties of the quasicrystal metasurfaces. (a,d) Measured wavelength dependent SHG responses of the Penrose and the HQC asymmetric samples. The SHG peak intensities with the same and opposite spin polarization states relative to that of fundamental wave (FW) are measured. The SHG signal for opposite spin states (LCP-RCP, RCP-LCP) are much stronger than for the same ones (LCP-LCP, RCP-RCP). The peak responses are located at wavelengths of 1275 nm for the Penrose metasurface (a) and 1250 nm for HQC asymmetric metasurface (d). Inset images show basic decagon and dodecagon decorated with C3 meta-atoms. (b, e) Measured power dependencies of SHG on the metasurfaces in (a) and (d) at their resonant wavelengths (axes shown in logarithmic scale). The polarization measurement scheme is LCP(FW)-RCP(SHG). The values of the slope given by linear regression of the experimental data are close to 2.0, which are in agreement with SHG process. (c, f) The spectra of SHG waves from metasurfaces in (a) and (d). The SHG spectra are recorded at the fundamental wavelengths of 1275 nm and 1250 nm, respectively.

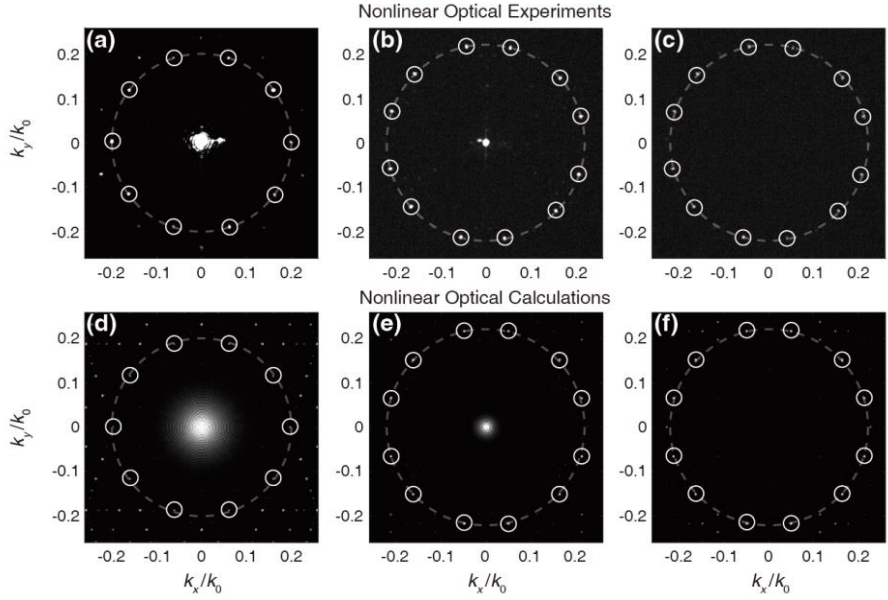

FIG. 4. Nonlinear optical diffractions of the quasicrystal metasurfaces. Three columns correspond to Penrose tiling, asymmetric and symmetric HQC tiling, respectively. (a-c), Measured and (d-f), calculated nonlinear optical diffraction patterns of quasicrystal metasurfaces [Fig. 2(a-c)]. The nonlinear optical diffraction patterns are recorded under LCP(FW)-RCP(SHG) polarization combination. The fundamental wavelengths are 1275 nm for Penrose quasicrystal (a) and 1250 nm for hexagonal quasicrystal designs (b and c). The calculated results are given by using far field dipole radiation model.